\documentclass[11pt]{article}
\usepackage[margin=1in]{geometry}
\usepackage{amsmath,amssymb,amsthm}
\usepackage{booktabs}
\usepackage{graphicx}
\usepackage{setspace}
\usepackage{enumitem}
\usepackage{xcolor}
\usepackage{algorithm}
\usepackage{algpseudocode}
\usepackage[round,authoryear]{natbib}
\usepackage[colorlinks=true,linkcolor=blue,citecolor=blue,urlcolor=blue]{hyperref}
\usepackage{xurl}

\newtheorem{proposition}{Proposition}
\newtheorem{theorem}{Theorem}
\newtheorem{corollary}{Corollary}
\newtheorem{lemma}{Lemma}
\theoremstyle{definition}
\newtheorem{definition}{Definition}
\newtheorem{assumption}{Assumption}
\theoremstyle{remark}
\newtheorem{remark}{Remark}

\newcommand{\E}{\mathbb{E}}
\newcommand{\Prob}{\mathbb{P}}
\newcommand{\indep}{\perp\!\!\!\perp}
\newcommand{\Sset}{\mathcal{S}}
\newcommand{\Tset}{\mathcal{T}}
\newcommand{\Rset}{\mathcal{R}}
\newcommand{\Hist}{\bar H}
\newcommand{\bara}{\bar a}
\newcommand{\pin}{\psi_{\mathrm{in}}}
\newcommand{\pout}{\psi_{\mathrm{out}}}
\newcommand{\ind}{\mathbb 1}

\title{Where the Data End and the Mechanism Begins: Doubly Robust Causal Digital Twins for Treatment Strategies}

\author{
Cong Cao\\
Department of Surgery, Stanford University School of Medicine,\\ Stanford, CA 94305, USA\\
Department of Biostatistics, Yale School of Public Health, New Haven, CT 06520, USA\\
Corresponding author: Cong Cao\\
Email: \texttt{congc@stanford.edu}
}
\date{}

\begin{document}

\maketitle

\begin{abstract}
Digital twins are increasingly used to ask how a patient's health would evolve under alternative treatment strategies. When a twin couples a mechanistic model with learned components and is fitted to observational records, its answer mixes information from the data with assumptions from the mechanism. For sustained strategies that begin at a clinical landmark, we split the counterfactual mean into a part the data support, where patients follow the strategy with probability bounded away from zero, and a part they do not. On the supported part, a cross-fitted sequential doubly robust estimator that uses the twin as its outcome model is consistent whenever the decision model is, so twins that fit equally well but disagree as plug-in estimators agree after correction. On the unsupported part, we characterize the values compatible with the data and with a residual of size at most $\epsilon$ around the mechanism. This set is nested in $\epsilon$, reduces to a point when the mechanism is exact and identifiable, and recovers assumption-free bounds as $\epsilon$ grows, so the penalty weight of a physics-informed loss becomes a sensitivity parameter. Knowledge constraints on how covariates enter the mechanism shrink the set, and when support grows over time, back-testing gives a data-based lower bound on $\epsilon$. In simulations based on an energy-balance model of weight change under glucagon-like peptide-1 receptor agonist therapy, the correction reduced the spread of estimates across four twins from 0.66 to 0.03 percentage points of weight loss with conservative coverage; on unsupported patients, the exact mechanism was rejected in every replicate, the bounds covered the truth at every residual size not rejected, and a mechanism-free extrapolation was biased by 0.5 points.
\end{abstract}

\noindent\textbf{Keywords:} causal inference; digital twins; physics-informed machine learning; hybrid mechanistic models; time-varying confounding; sequential double robustness; limited overlap; partial identification; sensitivity analysis

\section{Introduction}
\label{sec:intro}

About half of patients who start a glucagon-like peptide-1 receptor agonist (GLP-1RA) for obesity stop within the first year \citep{west2026,budini2026}. In randomized withdrawal trials, most of the weight lost is regained within a year of stopping \citep{rubino2021step4,aronne2024surmount4,wilding2022step1ext}, and a nonlinear meta-regression of six trials estimated that 60\% of lost weight returns at one year, with regain approaching a plateau at roughly three quarters of the loss \citep{budini2026}. In routine care, the average regain after discontinuation is much smaller, in part because many patients restart the drug or switch to another treatment \citep{gasoyan2026}. A clinician who reviews a patient six months after initiation therefore faces a question that neither source answers directly: for this patient, what would happen to body weight under continuation, under discontinuation, and under a switch, and how much of that answer can be trusted?

Digital twins are designed to answer such questions by simulating a patient's trajectory under alternative interventions \citep{bjornsson2020,nasem2024}. In physiology, the most credible twins combine a mechanistic model with learned components, for example an energy-balance model of weight change \citep{hall2011,hall2024plateau} augmented by a neural network that captures what the equations miss \citep{raissi2019,yin2021aphynity}. Fitted to observational records, such a twin can be run forward under any strategy. The difficulty is that it returns an answer whether or not the data contain patients who followed that strategy. Where they do, the answer can be checked against what happened to comparable patients; where they do not, it is an extrapolation carried entirely by the mechanism and by the inductive biases of the learned component. The two kinds of answer look identical in the output.

Recent results show why this matters. Hybrid models that fit the data well can lose the causal validity of their mechanistic core as their learned components become more flexible \citep{zou2024hybrid2}. Learned transport models with nearly identical predictive accuracy induced exposure mappings that implied different spillover effects \citep{cao2026interference}, and low prediction error of nuisance functions did not identify the better causal estimator \citep{cao2026prediction}. Observational data alone cannot certify that a twin is causally correct without assumptions such as no unmeasured confounding \citep{cornish2026}. Hybrid mechanistic models improve counterfactual prediction out of distribution \citep{meir2026structured,holt2024hdtwin}, and partial mechanistic knowledge can support what-if predictions without controlled experiments under explicit structural assumptions \citep{reroute2026}, but these approaches target prediction accuracy and do not separate what the data identify from what the mechanism supplies. Neural methods for counterfactual outcomes over time handle time-varying confounding \citep{seedat2022tecde,li2021gnet,frauen2023deepace} and quantify predictive uncertainty \citep{hess2024bncde}, but they contain no mechanism and therefore no principled basis for extrapolation. Debiased estimation has been extended to functionals of neural-operator trajectories \citep{hess2026dope}, without treatment strategies or confounding. In the literature on limited overlap, effects in regions without support are either excluded by trimming \citep{crump2009}, extrapolated by flexible regression \citep{nethery2019}, or bounded under smoothness assumptions on the outcome function \citep{ma2025overlap}; none uses a mechanism as the basis for extrapolation.

This paper makes four contributions. First, for sustained strategies that begin at a landmark, we decompose the counterfactual mean into a supported part, which is nonparametrically identified, and an unsupported part, which is not (Section~\ref{sec:decomp}). Second, we show that a cross-fitted sequential doubly robust estimator that uses the twin as its outcome model is consistent for the supported part whenever the decision model is, and we derive its exact second-order remainder; as a consequence, twins that are observationally equivalent but disagree as plug-in estimators agree after correction (Section~\ref{sec:dr}). Third, for the unsupported part we characterize the set of values compatible with the data and with a residual of size at most $\epsilon$ around the mechanism, establish its monotonicity, validity, and limiting behaviour, and show that knowledge constraints on the parameter encoder shrink it (Section~\ref{sec:bounds}); because a physics-informed penalty weight controls the residual size, it becomes a sensitivity parameter with a causal interpretation. Fourth, when the set of supported histories grows over calendar time, we show that back-testing yields a lower bound on the residual size required by the data and falsifies smaller values (Section~\ref{sec:calibration}). Section~\ref{sec:simulation} specifies and reports a simulation study based on published physiology and trial results.

We do not address the choice of adjustment set. Throughout, the adjustment set is the observed history itself. When the history is compressed into a learned representation, the requirements of \citet{cao2026reconstruction} apply in addition to those developed here; in particular, representation bias is first order and is not removed by the corrections below. The mechanism enters this paper only through the model for outcome dynamics.

\section{Setup and estimands}
\label{sec:setup}

\subsection{Data structure}

Time zero is treatment initiation. At a landmark $\tau$, for example six months, and at subsequent decision times $\tau=t_0<t_1<\cdots<t_{K-1}<t_K=T$, a treatment status $A_k\in\mathcal A$ is recorded, where in the running example $\mathcal A=\{\mathrm{c},\mathrm{s},\mathrm{w}\}$ denotes continuing the current GLP-1RA, stopping, or switching to a different agent. We observe $n$ independent copies of
\[
O=(H_0,A_0,L_1,A_1,\ldots,L_{K-1},A_{K-1},Y),
\]
where $H_0$ is the landmark history, consisting of baseline covariates $X$, the weight trajectory before and after initiation up to $\tau$, dose changes, and symptoms; $L_k$ collects the information recorded in $(t_{k-1},t_k]$, including weights measured at irregular visits and symptoms; and $Y$ is the outcome at $T$, for example the percentage of total weight loss relative to the weight at initiation. Write $\Hist_k=(H_0,A_0,L_1,\ldots,A_{k-1},L_k)$ for the history available at $t_k$, with $\Hist_0=H_0$.

\subsection{Strategies and estimands}

A sustained strategy $\bara=(a,\ldots,a)$ assigns treatment status $a$ at every decision time from the landmark onwards; for $a=\mathrm{s}$ it means stopping at $\tau$ and not restarting before $T$. Let $Y^{\bara}$ denote the potential outcome under $\bara$. The target is the counterfactual mean $\psi(a)=\E(Y^{\bara})$ and contrasts such as $\psi(\mathrm s)-\psi(\mathrm c)$. Strategies are defined at the landmark, as in target trial emulation \citep{hernan2016}; a patient who stops and later restarts has deviated from $\bar{\mathrm s}$, and the restart is part of the decision process rather than of the strategy. This distinction is what separates the strategy estimand from the natural course after discontinuation reported in routine-care studies \citep{gasoyan2026}.

For fixed $a$, define the decision probabilities and the iterated conditional expectations
\[
\pi_k(\bar h_k)=\Prob(A_k=a\mid \Hist_k=\bar h_k),\qquad
Q_K=Y,\qquad Q_k(\bar h_k)=\E\{Q_{k+1}\mid \Hist_k=\bar h_k, A_k=a\},
\]
for $k=K-1,\ldots,0$, where $Q_{k+1}$ is evaluated at $\Hist_{k+1}$ with $A_k=a$. Each $Q_k$ is a functional of the observed-data law wherever $\pi_k>0$.

\subsection{Identification and partial positivity}
\label{sec:decomp}

\begin{assumption}[Sequential identification]\label{ass:seq}
(i) Consistency: $Y=Y^{\bara}$ and $L_k=L_k^{\bara}$ when $\bar A=\bara$. (ii) Sequential exchangeability: $Y^{\bara}\indep A_k\mid \Hist_k,\ \bar A_{k-1}=\bar a_{k-1}$ for $k=0,\ldots,K-1$.
\end{assumption}

We do not assume positivity everywhere. Call a history $\bar h_k$ reachable from $h_0$ under $\bara$ if it has positive density under the law obtained by setting $A_j=a$ for $j<k$.

\begin{assumption}[Overlap gap]\label{ass:gap}
There is $\delta>0$ such that, for every $k$ and every history reachable under $\bara$, $\pi_k(\bar h_k)\in\{0\}\cup[\delta,1]$. Let $\Sset$ be the set of landmark histories $h_0$ from which $\pi_k\ge\delta$ holds for every reachable continuation and every $k$, and let $p_{\Sset}=\Prob(H_0\in\Sset)\in(0,1)$.
\end{assumption}

The gap is a simplification. In practice $\pi_k$ takes small positive values, and a threshold $\delta$ is chosen so that histories with smaller probabilities are treated as unsupported, trading the variance of weighting against reliance on the mechanism (Remark~\ref{rem:threshold}). Histories fail to be supported for clinical reasons: a patient with a large early weight loss and no side effects may essentially never stop at six months, so that the strategy $\bar{\mathrm s}$ is unsupported for that patient type even though such patients are common.

Define the supported and unsupported counterfactual means
\[
\pin(a)=\E\{Y^{\bara}\mid H_0\in\Sset\},\qquad \pout(a)=\E\{Y^{\bara}\mid H_0\notin\Sset\}.
\]

\begin{proposition}[Decomposition and identification]\label{prop:decomp}
Under Assumptions~\ref{ass:seq} and~\ref{ass:gap}:
\begin{enumerate}[label=(\alph*),leftmargin=*,itemsep=1pt]
\item $\psi(a)=p_{\Sset}\,\pin(a)+(1-p_{\Sset})\,\pout(a)$.
\item $\pin(a)=\E\{Q_0(H_0)\mid H_0\in\Sset\}$, so the supported part is identified.
\item If $\pi_0(h_0)=0$ for every $h_0\notin\Sset$ and $Y\in[y_{\mathrm{lo}},y_{\mathrm{hi}}]$, the sharp identified set for $\pout(a)$ under Assumptions~\ref{ass:seq} and~\ref{ass:gap} is $[y_{\mathrm{lo}},y_{\mathrm{hi}}]$.
\end{enumerate}
\end{proposition}

Part (b) is the g-formula restricted to supported histories \citep{robins1986}. Part (c) states that the observed data place no constraint on $\pout$ beyond the outcome range when the strategy is never initiated from the unsupported landmark histories, which is the situation in the running example and in the simulation. When positivity fails only at later decision times, the data carry information up to that time and the identified set is narrower; we do not characterize it here. Any informative statement about $\pout$ must therefore come from assumptions, and the question is how to state them so that their strength is explicit.

\section{Mechanism-informed causal digital twins}
\label{sec:twin}

\subsection{Mechanistic core}

Let $x(t)$ denote weight change from initiation and $u_a(t)\in[0,1]$ the drug exposure implied by strategy $a$ after the landmark: $u_{\mathrm c}(t)=1$; $u_{\mathrm s}(t)=\exp\{-(t-\tau)/\tau_d\}$ for $t\ge\tau$, with washout time constant $\tau_d$; and $u_{\mathrm w}$ the exposure to the new agent. We use the linearized energy-balance model of \citet{hall2011} with the feedback control of intake described by \citet{hall2024plateau}. Changes in energy intake and expenditure are
\[
\Delta I(t)=-P\,u(t)-k\{u(t)\}\,x(t),\qquad \Delta E(t)=\beta_E\,x(t),
\]
where $P$ is the persistent drug-induced reduction in intake, $k(u)=k^{\mathrm{off}}-(k^{\mathrm{off}}-k^{\mathrm{on}})\,u$ is the strength of the feedback that raises appetite as weight falls, and $\beta_E$ is the change in expenditure per unit weight change. With energy density $\rho$, the mechanistic drift is
\begin{equation}\label{eq:mech}
f(x,u;\theta)=-\frac{1}{\rho}\big[P\,u+\{\beta_E+k(u)\}\,x\big],\qquad \theta=(P,k^{\mathrm{on}},k^{\mathrm{off}}),
\end{equation}
with drug-specific values of $(P,k^{\mathrm{on}})$ when $\mathcal A$ contains more than one agent. Group-level fits of this model indicate that GLP-1RAs and bariatric surgery weaken the appetite feedback by roughly 40--70\% relative to its baseline strength \citep{hall2024plateau}.

\subsection{Twin}

\begin{definition}[Twin]\label{def:twin}
A twin $\mathsf T=(g,r)$ consists of an encoder $g$ that maps baseline covariates to mechanistic parameters, $\theta_i=g(X_i)$, and a residual $r(x,u,z,t)$, where $z$ collects features of the landmark history. Under strategy $a$, a twin generates post-landmark weight change for patient $i$ by
\[
dx_i(t)=\big[f\{x_i(t),u_a(t);g(X_i)\}+r\{x_i(t),u_a(t),z_i,t\}\big]\,dt+\sigma\,dB_i(t),\qquad t\ge\tau,
\]
started from the state at $\tau$, with weights observed with measurement error at visit times. Under the decision process observed in the data, the same equation with $u$ given by the observed treatment path defines the twin's law of the observed trajectories. A twin implies functions $Q_k^{\mathsf T}(\bar h_k)$, obtained by conditioning the latent state on the weights observed up to $t_k$ and simulating forward under $\bara$.
\end{definition}

The same twin also describes the pre-treatment window with $u=0$, where the feedback strength is $k^{\mathrm{off}}$. This is one route by which the mechanism carries information across strategies (Remark~\ref{rem:info}).

\paragraph{Knowledge constraints.} Prior knowledge can restrict which covariates may affect which parameter. A mask $M=(\mathcal J_1,\ldots,\mathcal J_q)$ requires the $j$th component of $g$ to depend on $X$ only through $X_{\mathcal J_j}$; for example, the drug effect $P$ may depend on glycaemic status and genetic risk for type 2 diabetes, whereas the off-drug feedback $k^{\mathrm{off}}$ may depend on demographic characteristics and the pre-treatment weight trajectory. Let $\mathcal G_M$ denote encoders that respect $M$.

\paragraph{Residual size.} For a reference measure $\mu$ on $(x,u,z,t)$, let $\|r\|_\mu^2=\int r^2\,d\mu$ and $\Rset_\epsilon=\{r\in\Rset:\|r\|_\mu\le\epsilon\}$ for a function class $\Rset$. The measure $\mu$ must cover the states, exposures, and history features at which extrapolation is required, not only those seen in the data; otherwise the residual is unconstrained exactly where the answer depends on it. In the simulation, $r$ acts only off treatment and is linear in features, $r=(1-u)(b_0+b^\top z+b_x x_\tau)$, where $x_\tau$ is the observed weight change at the landmark, and $\mu$ is the product of the empirical law of $(z,x_\tau)$ over all patients, supported or not, and the uniform law of $u$ on $[0,1]$; then $\|r\|_\mu^2=\tfrac13\,b^\top\E(vv^\top)b$ with $v=(1,z,x_\tau)$.

\begin{remark}[Physics-informed penalties and residual size]\label{rem:lambda}
Physics-informed training minimizes a penalized loss $\mathcal L_n(\mathsf T)+\lambda\|r\|_\mu^2$, where $\mathcal L_n$ measures fit to the observed trajectories \citep{raissi2019,cao2025pbml}; APHYNITY penalizes the norm of the learned component for a related purpose \citep{yin2021aphynity}. When the problem is convex in $r$, Lagrangian duality gives a decreasing map $\epsilon\mapsto\lambda(\epsilon)$ under which the penalized and constrained problems $\min\{\mathcal L_n(\mathsf T):\|r\|_\mu\le\epsilon\}$ share solutions. For neural residuals the correspondence holds only locally, so we index all results by the realized residual size rather than by $\lambda$. Either way, the penalty weight sets how far the twin may depart from the mechanism, which Section~\ref{sec:bounds} turns into a statement about counterfactuals.
\end{remark}

\begin{definition}[Observationally equivalent twins]\label{def:equiv}
A twin is compatible with the observed-data law $P$ if, for every $k$ and every action with positive probability given the history, the twin reproduces the conditional law of $L_{k+1}$ given $(\Hist_k,A_k)$ and of $Y$ given $(\Hist_{K-1},A_{K-1})$, and reproduces the pre-treatment trajectories. Let
\[
\Tset(\epsilon,M)=\{\mathsf T=(g,r):\ g\in\mathcal G_M,\ r\in\Rset_\epsilon,\ \mathsf T\ \text{compatible with}\ P\}.
\]
\end{definition}

Compatibility uses all observed treatment paths, not only those consistent with $\bara$. Because the mechanistic parameters are shared across strategies, patients who continued treatment constrain $P$ and $k^{\mathrm{on}}$, and the pre-treatment window constrains $k^{\mathrm{off}}$, even for patient types who never stop.

\begin{remark}[Where the information for extrapolation comes from]\label{rem:info}
Under~\eqref{eq:mech}, the trajectory after stopping is governed by the decay of $P\,u(t)$ and by the return of the feedback from $k^{\mathrm{on}}$ to $k^{\mathrm{off}}$. For a patient who never stops, $P$ and $k^{\mathrm{on}}$ are informed by the on-treatment trajectory and $k^{\mathrm{off}}$ by the pre-treatment trajectory and by the encoder, which is learned from patients with similar covariates who did stop. Strategy positivity fails because the decision depends on early response, a post-baseline quantity, while the covariates $X$ that enter the encoder may overlap well between supported and unsupported patients. The mechanism is what links the two.
\end{remark}

\section{The supported part: doubly robust twins}
\label{sec:dr}

Fix $\Sset$; in practice it is estimated by thresholding estimated decision probabilities on a split of the data that is not used for estimation. Divide the estimation sample into $J$ folds. For observations in fold $j$, let $\hat\pi_k$ and a twin $\hat{\mathsf T}$ be fitted on the other folds, set $\hat Q_k=Q_k^{\hat{\mathsf T}}$ and $\hat Q_K=Y$, and define the weights $\hat W_{-1}=1$ and $\hat W_k=\hat W_{k-1}\ind(A_k=a)/\hat\pi_k(\Hist_k)$. The estimator of the supported part is
\begin{equation}\label{eq:dr}
\hat\psi_{\mathrm{in}}(a)=\frac{\sum_{i=1}^n\ind(H_{0i}\in\Sset)\,\hat\varphi(O_i)}{\sum_{i=1}^n\ind(H_{0i}\in\Sset)},\qquad
\hat\varphi(O)=\hat Q_0(H_0)+\sum_{k=0}^{K-1}\hat W_k\big\{\hat Q_{k+1}(\Hist_{k+1})-\hat Q_k(\Hist_k)\big\}.
\end{equation}
This is the sequential augmented inverse-probability-weighted estimator \citep{bang2005,vanderlaan2012ltmle,luedtke2017sequential} with the iterated regressions supplied by the twin rather than by separate regression fits. The corresponding influence function is
\[
D_{\mathrm{in}}(O)=\frac{\ind(H_0\in\Sset)}{p_{\Sset}}\Big[\sum_{k=0}^{K-1}W_k\{Q_{k+1}(\Hist_{k+1})-Q_k(\Hist_k)\}+Q_0(H_0)-\pin(a)\Big],
\]
with $W_k$ the weights at the true decision probabilities.

\begin{lemma}[Exact remainder]\label{lem:remainder}
For fixed $\hat\pi_k$ bounded below on supported histories and fixed $\hat Q_k$,
\begin{multline*}
\E\{\ind(H_0\in\Sset)\hat\varphi(O)\}-\E\{\ind(H_0\in\Sset)Q_0(H_0)\}\\
=\sum_{k=0}^{K-1}\E\Big[\ind(H_0\in\Sset)\,\hat W_{k-1}\,\frac{\hat\pi_k(\Hist_k)-\pi_k(\Hist_k)}{\hat\pi_k(\Hist_k)}\,\big\{\hat Q_k(\Hist_k)-Q_k(\Hist_k)\big\}\Big].
\end{multline*}
\end{lemma}

Each term is the product of the error in one decision model and the error in the twin's regression at the same time point, so the bias vanishes if, at every $k$, either the decision model or the twin is correct.

\begin{theorem}[Doubly robust twins on the supported part]\label{thm:dr}
Suppose Assumptions~\ref{ass:seq} and~\ref{ass:gap} hold, $\hat\pi_k\ge\delta/2$ on supported histories with probability tending to one, and $\hat\pi_k\to\bar\pi_k$ and $\hat Q_k\to\bar Q_k$ in $L_2(P)$ on supported histories for some limits $\bar\pi_k$ and $\bar Q_k$.
\begin{enumerate}[label=(\alph*),leftmargin=*,itemsep=1pt]
\item If, for each $k$, either $\bar\pi_k=\pi_k$ or $\bar Q_k=Q_k$, then $\hat\psi_{\mathrm{in}}(a)\to\pin(a)$ in probability.
\item If $\bar\pi_k=\pi_k$ and $\bar Q_k=Q_k$ for all $k$ and $\sum_{k}\|\hat\pi_k-\pi_k\|\,\|\hat Q_k-Q_k\|=o_P(n^{-1/2})$, with norms over supported histories, then $\sqrt n\{\hat\psi_{\mathrm{in}}(a)-\pin(a)\}\to N\{0,\mathrm{Var}(D_{\mathrm{in}})\}$.
\item Let $\hat{\mathsf T}^{(1)}$ and $\hat{\mathsf T}^{(2)}$ be two twin-fitting procedures. If both satisfy the conditions of (b), then $\hat\psi_{\mathrm{in}}^{(1)}(a)-\hat\psi_{\mathrm{in}}^{(2)}(a)=o_P(n^{-1/2})$. If both are misspecified, with limits $\bar Q^{(1)}\ne\bar Q^{(2)}$, but the decision models are consistent, both estimators converge to $\pin(a)$, whereas the plug-in estimators converge to $\E\{\bar Q_0^{(m)}(H_0)\mid H_0\in\Sset\}$, $m=1,2$, which differ in general.
\end{enumerate}
\end{theorem}

Part (c) is the formal version of an empirical phenomenon: twins that fit the observed trajectories equally well, and that would pass any check based on prediction error \citep{cao2026prediction}, can imply different counterfactuals, as learned transport models did in \citet{cao2026interference}. On supported histories the correction removes this disagreement to first order. Under misspecification of the twin, consistency of $\hat\psi_{\mathrm{in}}$ rests on the decision models, and the rate is that of $\hat\pi_k$; with correctly specified parametric decision models it is $\sqrt n$.

\begin{remark}[Choice of threshold]\label{rem:threshold}
A larger $\delta$ shrinks $\Sset$, stabilizes the weights, and moves more of $\psi(a)$ into the part carried by the mechanism. Reporting $\hat\psi_{\mathrm{in}}$, the bounds of Section~\ref{sec:bounds}, and $p_{\Sset}$ over a grid of $\delta$ makes this trade-off visible. We do not study asymptotics with $\delta_n\to0$.
\end{remark}

\section{The unsupported part: mechanism-indexed bounds}
\label{sec:bounds}

For a twin $\mathsf T$, let $\pout(\mathsf T)=\E\{Q_0^{\mathsf T}(H_0)\mid H_0\notin\Sset\}$ be the value it implies on unsupported histories, and define
\[
B(\epsilon,M)=\Big[\inf_{\mathsf T\in\Tset(\epsilon,M)}\pout(\mathsf T),\ \sup_{\mathsf T\in\Tset(\epsilon,M)}\pout(\mathsf T)\Big].
\]

\begin{assumption}[Mechanism adequacy at level $\epsilon^*$]\label{ass:adequacy}
The potential trajectories under every strategy are generated, given the landmark history, by a twin $\mathsf T^*=(g^*,r^*)$ with $g^*\in\mathcal G_{M^*}$ and $\|r^*\|_\mu\le\epsilon^*$.
\end{assumption}

The level $\epsilon^*$ is unknown. Assumption~\ref{ass:adequacy} with a small $\epsilon^*$ is a strong statement about physiology; with a large $\epsilon^*$ it is weak.

\begin{proposition}[Properties of the bounds]\label{prop:bounds}
Suppose Assumptions~\ref{ass:seq} and~\ref{ass:gap} hold.
\begin{enumerate}[label=(\alph*),leftmargin=*,itemsep=1pt]
\item \emph{Monotonicity.} If $\epsilon\le\epsilon'$ then $B(\epsilon,M)\subseteq B(\epsilon',M)$; if $\mathcal G_M\subseteq\mathcal G_{M'}$ then $B(\epsilon,M)\subseteq B(\epsilon,M')$.
\item \emph{Validity.} Under Assumption~\ref{ass:adequacy}, $\pout(a)\in B(\epsilon,M)$ for every $\epsilon\ge\epsilon^*$ and every $M$ with $\mathcal G_{M^*}\subseteq\mathcal G_M$.
\item \emph{Exact mechanism.} Suppose the parameters are identified at $\epsilon=0$: any two twins in $\Tset(0,M)$ have encoders that agree almost surely on the support of $X$ among supported patients and in the pre-treatment window, and the support of $X$ among unsupported patients is contained in that set. Then $B(0,M)$ is a single point if $\Tset(0,M)$ is non-empty. If $\Tset(0,M)$ is empty, the exact mechanism with mask $M$ is refuted by the observed data.
\item \emph{No mechanism.} Let $\mathcal I_0$ be the identified set for $\pout(a)$ under Assumptions~\ref{ass:seq} and~\ref{ass:gap} alone. Suppose the residual class is rich: for every $c\in\mathcal I_0$ and every $\eta>0$ there is a twin, compatible with $P$ and with finite residual size, whose out-of-support value lies within $\eta$ of $c$. Then $\bigcup_{\epsilon}B(\epsilon,M)$ is dense in $\mathcal I_0$, and $B(\epsilon,M)\subseteq\overline{\mathcal I_0}$ for every $\epsilon$.
\end{enumerate}
\end{proposition}

Part (a) says that a stronger trust in the mechanism (smaller $\epsilon$) and stronger knowledge about how covariates enter it (a smaller mask) both narrow the bounds. Part (b) says that the narrowing is valid only if the trust is warranted. Parts (c) and (d) give the two ends of the scale: point identification by the mechanism, and the assumption-free set of Proposition~\ref{prop:decomp}(c). Reporting $B(\epsilon,M)$ as a function of $\epsilon$ shows how strongly a conclusion about unsupported patients depends on the mechanism; this is the role that sensitivity parameters play in analyses of unmeasured confounding, here attached to the physics-informed penalty. Part (c) also supplies a falsification test: if no twin with small residual is compatible with the data, the mechanism with that residual size is rejected, in the spirit of \citet{cornish2026}.

Combining with Proposition~\ref{prop:decomp}(a), the counterfactual mean satisfies
\[
\psi(a)\in p_{\Sset}\,\pin(a)+(1-p_{\Sset})\,B(\epsilon,M)\quad\text{for all }\epsilon\ge\epsilon^*.
\]

\subsection{Computation and inference}
\label{sec:computation}

The endpoints of $B(\epsilon,M)$ are values of optimization problems over twins, as identified sets for counterfactuals in structural dynamic models are \citep{kalouptsidi2020}. We approximate them with extremal twins (Algorithm~\ref{alg:extremal}): for each sign, the twin's out-of-support value is pushed up or down subject to a fit constraint that admits near-minimizers of the empirical loss, in the spirit of set estimation with near-minimizers \citep{chernozhukov2007}. The fit threshold is set once, by the best fit in the richest residual class considered, so that the estimated sets are nested in $\epsilon$ as in Proposition~\ref{prop:bounds}(a); if no twin with residual size $\epsilon$ attains the threshold, the estimated set is empty and the mechanism with that residual size is rejected, the sample counterpart of Proposition~\ref{prop:bounds}(c). Confidence intervals for the endpoints can be obtained by the bootstrap and combined into an interval for the partially identified parameter \citep{imbens2004}.

In implementation, compatibility is imposed through the mean trajectory: a twin is compatible if it minimizes the population loss $\mathcal L(\theta)$, the expected squared error of predicted weights along the observed treatment paths, over the class $\Theta_\epsilon$ of twin parameters with residual size at most $\epsilon$. Write $\Theta_0(\epsilon)=\arg\min_{\Theta_\epsilon}\mathcal L$ and $\hat\Theta(\epsilon)=\{\theta\in\Theta_\epsilon:\mathcal L_n(\theta)\le\inf_{\Theta_\epsilon}\mathcal L_n+c_n\}$.

\begin{proposition}[Consistency of extremal twins]\label{prop:setcons}
Suppose that $\Theta_\epsilon$ is compact, that $\sup_{\Theta_\epsilon}|\mathcal L_n-\mathcal L|=O_P(a_n)$ with $a_n\to0$, that $\theta\mapsto\pout(\theta)$ is continuous on $\Theta_\epsilon$ and its sample analogue converges to it uniformly in probability, and that $c_n\to0$ with $c_n/a_n\to\infty$. Then $\hat\Theta(\epsilon)$ converges to $\Theta_0(\epsilon)$ in Hausdorff distance in probability, and the endpoints of $\{\hat\psi_{\mathrm{out}}(\theta):\theta\in\hat\Theta(\epsilon)\}$ converge in probability to $\min$ and $\max$ of $\pout$ over $\Theta_0(\epsilon)$.
\end{proposition}

Under Assumption~\ref{ass:adequacy}, the true twin minimizes $\mathcal L$ whenever $\epsilon\ge\epsilon^*$, so $\Theta_0(\epsilon)$ is the set of compatible twins in this mean sense, and the limits are the endpoints of $B(\epsilon,M)$. In the simulation, $a_n$ is of order $n^{-1/2}$ and we use the relative slack $c_n=2\sqrt{2/N}$, with $N$ the number of weight measurements.

\begin{algorithm}[htbp]
\caption{Extremal twins for $B(\epsilon,M)$}
\label{alg:extremal}
\begin{algorithmic}[1]
\Require increasing grid $\epsilon_1<\cdots<\epsilon_J$, mask $M$, relative slack $c_n>0$
\State Fit the richest twin $\hat{\mathsf T}_{\mathrm{rich}}$ and set the threshold $\tau_n=(1+c_n)\,\mathcal L_n(\hat{\mathsf T}_{\mathrm{rich}})$
\For{$j=1,\ldots,J$}
  \State $\hat{\mathsf T}_j=\arg\min\{\mathcal L_n(\mathsf T): g\in\mathcal G_M,\ \|r\|_\mu\le\epsilon_j\}$
  \If{$\mathcal L_n(\hat{\mathsf T}_j)>\tau_n$} \State report $B(\epsilon_j,M)=\emptyset$ and continue \EndIf
  \For{$s\in\{-1,+1\}$}
    \State $\hat{\mathsf T}_{j,s}=\arg\max\{s\,\hat\psi_{\mathrm{out}}(\mathsf T): g\in\mathcal G_M,\ \|r\|_\mu\le\epsilon_j,\ \mathcal L_n(\mathsf T)\le\tau_n\}$, started at $\hat{\mathsf T}_j$
  \EndFor
  \State Record the endpoints, taking the envelope with those at $\epsilon_{j-1}$
\EndFor
\end{algorithmic}
\end{algorithm}

\section{Calibrating the residual size over time}
\label{sec:calibration}

The size $\epsilon$ is a sensitivity parameter, but data can sometimes say which values are too small. Treatment practice changes over calendar time, for example through drug shortages or changes in insurance coverage, and patient types for whom a strategy was unsupported in an early era may follow it in a later one.

Let $e\in\{1,2\}$ index two eras with observed-data laws $P^{(1)}$ and $P^{(2)}$ and supported sets $\Sset_1\subseteq\Sset_2$, and let $\mathcal V=\Sset_2\setminus\Sset_1$ be the back-test region, unsupported in era~1 and supported in era~2.

\begin{assumption}[Invariant dynamics]\label{ass:invariance}
The law of the potential trajectories under each strategy, given the landmark history, is the same in both eras. The eras may differ in the distribution of histories, in the decision process, and in the visit process.
\end{assumption}

Let $B^{(1)}_{\mathcal V}(\epsilon,M)$ be the bounds computed from twins compatible with $P^{(1)}$, with the out-of-support value averaged over the era-2 distribution of histories in $\mathcal V$, and let $\psi_{\mathcal V}=\E^{(2)}\{Q_0^{(2)}(H_0)\mid H_0\in\mathcal V\}$, which is identified in era~2. Define
\[
\epsilon_{\mathrm{cal}}=\inf\{\epsilon\ge0:\ \psi_{\mathcal V}\in B^{(1)}_{\mathcal V}(\epsilon,M)\}.
\]

\begin{proposition}[Back-testing]\label{prop:calibration}
Suppose Assumptions~\ref{ass:seq} and~\ref{ass:gap} hold in both eras and Assumptions~\ref{ass:adequacy} and~\ref{ass:invariance} hold, with $\mathcal G_{M^*}\subseteq\mathcal G_M$.
\begin{enumerate}[label=(\alph*),leftmargin=*,itemsep=1pt]
\item $\epsilon_{\mathrm{cal}}\le\epsilon^*$.
\item Every $\epsilon<\epsilon_{\mathrm{cal}}$ is inconsistent with Assumption~\ref{ass:adequacy} at level $\epsilon$.
\item If, in addition, extrapolation to the remaining unsupported set $\mathcal U=\Sset_2^{c}$ requires no larger residual than extrapolation to $\mathcal V$, in the sense that $\inf\{\epsilon:\pout^{\mathcal U}(a)\in B^{(2)}_{\mathcal U}(\epsilon,M)\}\le\epsilon_{\mathrm{cal}}$, then $\pout^{\mathcal U}(a)\in B^{(2)}_{\mathcal U}(\epsilon,M)$ for every $\epsilon>\epsilon_{\mathrm{cal}}$, and also at $\epsilon=\epsilon_{\mathrm{cal}}$ if the infimum is attained.
\end{enumerate}
\end{proposition}

Back-testing therefore gives a lower bound on the residual size that the data require, not an upper bound. Parts (a) and (b) need only invariance; part (c) needs an additional assumption that the back-test region is at least as hard to extrapolate to as the target region, which cannot be checked. In the simulation design below it holds by construction.

In practice $\psi_{\mathcal V}$ is estimated by~\eqref{eq:dr} in era~2 and the bounds by Algorithm~\ref{alg:extremal} in era~1. For each $\epsilon$ in a grid $\epsilon_1<\cdots<\epsilon_J$, consider the null hypothesis $H_0(\epsilon):\psi_{\mathcal V}\in B^{(1)}_{\mathcal V}(\epsilon,M)$, rejected when the Wald interval $\hat\psi_{\mathcal V}\pm z_{1-\alpha}\widehat{\mathrm{se}}$ does not intersect the estimated bounds, and let $\hat\epsilon_{\mathrm L}$ be the smallest grid value at which $H_0$ is not rejected.

\begin{corollary}[Lower confidence bound]\label{cor:lcb}
Under the conditions of Proposition~\ref{prop:calibration}, Theorem~\ref{thm:dr}(b) applied in era~2, and Proposition~\ref{prop:setcons} applied in era~1, let $\epsilon_{j^*}$ be the smallest grid value with $\epsilon_{j^*}\ge\epsilon_{\mathrm{cal}}$, and suppose $\psi_{\mathcal V}$ lies in the interior of $B^{(1)}_{\mathcal V}(\epsilon_{j^*},M)$. Then $\liminf_n\Prob(\hat\epsilon_{\mathrm L}\le\epsilon_{j^*})\ge1-\alpha$.
\end{corollary}

Combined with Proposition~\ref{prop:calibration}(a), $\hat\epsilon_{\mathrm L}$ is an asymptotic lower confidence bound for the grid level just above $\epsilon_{\mathrm{cal}}$, and hence for nothing larger than $\epsilon^*$ rounded up to the grid.

\section{Simulation study}
\label{sec:simulation}

The study follows the ADEMP structure \citep{morris2019}; the code is a single script described in Appendix~\ref{app:sim}. All results below use $\omega=1$, the setting in which the mechanism omits part of the dynamics.

\paragraph{Aims.} (i) To verify that, on supported histories, twins that disagree as plug-in estimators agree after correction (Theorem~\ref{thm:dr}); (ii) to evaluate the validity and width of $B(\epsilon,M)$ on unsupported histories as functions of $\epsilon$ and of the mask, and the rejection of too small $\epsilon$ (Proposition~\ref{prop:bounds}); (iii) to evaluate back-testing (Proposition~\ref{prop:calibration} and Corollary~\ref{cor:lcb}); and (iv) to compare with plug-in twins, the mechanistic core alone, and a mechanism-free g-computation estimator.

\paragraph{Data-generating mechanism.} Baseline covariates are age, sex (60\% female), body mass index, glycated haemoglobin, and a genetic risk score for type 2 diabetes, standardized except sex; weight at initiation is $\max\{70,100+15\cdot\text{BMI}_z\}$ kg. Gastrointestinal symptoms $S$ at the landmark are binary with probability $\mathrm{expit}(-0.8+0.3\cdot\text{female})$, and a cost shock $C$, recorded in the landmark history, has probability 0.10 in era~1 and 0.35 in era~2. Weight change follows the energy-balance model of Section~\ref{sec:twin} on a weekly grid. The energy density and the expenditure slope are sex-specific values of the linearized NIH Body Weight Planner model, $\rho=9{,}383$ and $9{,}916$ kcal/kg and $\beta_E=28$ and $24$ kcal/day per kg for men and women \citep{guo2018}. Individual parameters are
\begin{gather*}
P_i=P_0\exp(0.15\,\text{BMI}_i+0.10\,\text{female}_i-0.15\,\text{HbA1c}_i-0.10\,\text{PRS}_i+0.15\,S_i),\\
k^{\mathrm{off}}_i=95\exp(0.10\,\text{age}_i-0.08\,\text{female}_i),\qquad k^{\mathrm{on}}_i=(1-w_i)\,k^{\mathrm{off}}_i,\\
w_i=0.4+0.3\,\mathrm{expit}(0.8\,\text{female}_i-0.6\,\text{HbA1c}_i),
\end{gather*}
so that the baseline feedback strength matches the value estimated by \citet{hall2024plateau} and treatment weakens it by 40--70\%. These coefficients define the true mask $M^*$: $P$ depends on sex, BMI, HbA1c, genetic risk, and symptoms; $w$ on sex and HbA1c; and $k^{\mathrm{off}}$ on age and sex. Drug exposure is $u=1$ on treatment and decays with time constant $\tau_r$ after stopping. Variation in the parameters is explained by the covariates, so that the encoder can represent it; unexplained variation enters through the intake noise below. The design does not include random effects in the parameters, which would require updating individual parameters from early weights. We set $P_0=800$ kcal/day, which gives mean total weight loss of 10.3\% at 26 weeks and 14.4\% at 78 weeks under continuation, close to the 15\% reported at 68 weeks in STEP~1 \citep{wilding2021step1}, and $\tau_r=14$ weeks, which, together with the residual described next, gives a median regain of 59\% of the lost weight 52 weeks after stopping, close to the 60\% estimated by \citet{budini2026}. Intake fluctuates weekly with standard deviation 100 kcal/day, and weights are recorded at weeks 4, 8, 12, 16, 20, and 26 and every 8 weeks thereafter up to week 78, with measurement error of standard deviation 1 kg.

The true dynamics include a residual that the mechanistic core omits: after a patient first stops, intake stays lower by an amount proportional to the weight lost by the landmark, $r^*=\omega\gamma\,x_\tau(1-u)$ with $\gamma=40$ kcal/day per kg, a stand-in for behavioural changes learned during treatment. Without it, the core predicts that all lost weight is eventually regained; with it, regain levels off below the pre-treatment weight, as in the meta-regression of \citet{budini2026}. This residual lies in the twin's residual class, so Assumption~\ref{ass:adequacy} holds with $\epsilon^*=\|r^*\|_\mu$, about 250 kcal/day and computed in each replicate. The dial $\omega\in\{0,1\}$ makes the mechanism exact ($\omega=0$) or not ($\omega=1$).

At the landmark $\tau=26$ weeks, the probability of stopping is $\mathrm{expit}\{-1.5+1.2S+1.0C-0.15(\text{early loss}-8)\}$, with early loss the observed percentage weight loss. In era~1 it is zero for patients with early loss of at least 12\% and no symptoms, who make up 10\% of the cohort and form the unsupported set for the stopping strategy. In era~2, a cost shock triggers stopping with probability 0.5 among these patients, so that those with a cost shock form the back-test region $\mathcal V$ and those without one remain unsupported. Because the cost shock is independent of the dynamics, the condition of Proposition~\ref{prop:calibration}(c) holds in this design. Every 8 weeks after the landmark, patients on treatment stop with probability $\mathrm{expit}(-3+C+0.5S)$, and patients off treatment restart with probability $\min\{0.9,\mathrm{expit}(-3.5+0.15\,\text{regain})\}$, where regain is the weight regained in kilograms since stopping. Restarting in response to regain creates time-varying confounding, and its frequency is comparable to the roughly 20\% of patients who restarted within a year in routine care \citep{gasoyan2026}. Each era has $n=2{,}000$ patients and each setting uses 200 replicates.

\paragraph{Estimands.} The counterfactual means of the percentage of total weight loss at 78 weeks under continuation, $\psi(\mathrm c)$, and under stopping at the landmark without restarting, $\psi(\mathrm s)$, with the supported and unsupported parts of $\psi(\mathrm s)$. True values are computed by simulating $60{,}000$ patients under each strategy.

\paragraph{Methods.} The twin has log-linear and logit-linear encoders for $P$, $w$, and $k^{\mathrm{off}}$, an estimated recovery time constant, and the residual $(1-u)(b_0+b^\top z+b_x x_\tau)$ of Section~\ref{sec:twin}, with $z$ the five covariates and $S$. Decision models are logistic regressions. The methods are: (M1) the mechanistic core without residual, as a plug-in estimator and with the correction~\eqref{eq:dr}; (M2) hybrid twins with penalty weights $\lambda\in\{10^{-7},10^{-5}\}$ and an unrestricted encoder, and (M3) a hybrid twin with the true mask, each as a plug-in estimator and with the correction; (M4) iterated linear regressions among strategy followers, a mechanism-free g-computation estimator that extrapolates linearly to unsupported patients; and (M5) the bounds $B(\epsilon,M)$ of Algorithm~\ref{alg:extremal} on unsupported histories, with unrestricted and true masks, over $\epsilon\in\{0,0.1,0.25,0.5,1,2\}\times\epsilon^*$. Corrections use two-fold cross-fitting with the same folds for all twins. Back-testing compares the era-2 corrected estimate on $\mathcal V$ with era-1 bounds evaluated on the era-2 histories in $\mathcal V$, and reports $\hat\epsilon_{\mathrm L}$ of Corollary~\ref{cor:lcb} with $\alpha=0.025$.

\paragraph{Performance measures.} On supported histories: bias, mean estimated standard error, and coverage of 95\% intervals for each twin, with and without correction, and the standard deviation across twins within a replicate before and after correction. On unsupported histories: coverage of $\pout(\mathrm s)$ by $B(\epsilon,M)$, its width, and the share of replicates in which the set is empty. For back-testing: the distribution of $\hat\epsilon_{\mathrm L}/\epsilon^*$ and the share of replicates with $\hat\epsilon_{\mathrm L}\le\epsilon^*$. Monte Carlo standard errors are given in the table notes.

\subsection{Results}
\label{sec:simresults}

All results are based on 200 replicates with $n=2{,}000$ patients per era. On average, 10.6\% of patients were unsupported for the stopping strategy, and the true residual size was $\epsilon^*\approx249$ kcal/day.

\paragraph{Supported histories.} Table~\ref{tab:supported} shows that the four twins, all fitted to the same data, gave plug-in estimates whose biases ranged from $-0.06$ to $-1.37$ percentage points of total weight loss. The mechanistic core and the strongly penalized hybrid were most biased because they cannot represent the intake reduction retained after stopping and therefore predict too much regain. After the correction~\eqref{eq:dr}, all biases were at most 0.04 points and the within-replicate standard deviation across twins fell from 0.656 to 0.029, the behaviour described in Theorem~\ref{thm:dr}(c). The two misspecified twins kept a small bias (0.042 and 0.030, about a third of their standard errors), as expected when consistency rests on the decision models alone (Theorem~\ref{thm:dr}(a)). Coverage of nominal 95\% intervals was 0.975 to 0.995, so the variance estimator was conservative; in the first 50 replicates, the empirical standard deviations of the corrected estimates were 10--20\% below the mean estimated standard errors. This is consistent with the known effect of treating estimated weights as known in inverse-probability-weighted estimators \citep{lunceford2004}. For continuation, which is supported for every patient, the corrected estimate had bias $-0.014$ and coverage 0.950. The mechanism-free iterated regression estimator was also nearly unbiased on supported histories (bias $-0.056$): where the data support a strategy, the mechanism is not needed.

\begin{table}[htbp]
\centering
\caption{Supported histories, strategy ``stop at the landmark'': bias of plug-in and corrected estimates of $\pin(\mathrm s)$ (true value 4.03 percentage points of total weight loss at 78 weeks), $\omega=1$, 200 replicates.}
\label{tab:supported}
\small
\begin{tabular}{lcccc}
\toprule
Twin & Plug-in bias & Corrected bias & Mean SE & Coverage \\
\midrule
Mechanistic core, no residual & $-1.369$ & 0.042 & 0.116 & 0.995 \\
Hybrid, $\lambda=10^{-7}$ & $-0.076$ & 0.005 & 0.075 & 0.975 \\
Hybrid, $\lambda=10^{-5}$ & $-0.969$ & 0.030 & 0.099 & 0.990 \\
Hybrid with true mask, $\lambda=10^{-7}$ & $-0.064$ & 0.006 & 0.075 & 0.980 \\
\midrule
SD across the four twins & 0.656 & 0.029 & & \\
\bottomrule
\end{tabular}

\smallskip
\parbox{0.92\linewidth}{\footnotesize SE, estimated standard error; SD, mean within-replicate standard deviation of the four estimates. Monte Carlo standard errors are about 0.005--0.008 for the corrected biases and 0.005--0.011 for the coverages.}
\end{table}

\paragraph{Unsupported histories.} The true value on unsupported histories was 5.67 points, and the mechanism-free estimator, which extrapolates linearly, was biased by $-0.49$ points while reporting a single number. Table~\ref{tab:bounds} shows the bounds. For $\epsilon\le0.25\,\epsilon^*$, no twin met the fit threshold in any replicate, so the estimated set was empty: the data rejected the exact mechanism and every residual smaller than a quarter of the true one, the sample counterpart of Proposition~\ref{prop:bounds}(c). For every residual size that was not rejected, the bounds covered the true value in at least 99.5\% of replicates. Their width increased with $\epsilon$, from 2.23 to 3.02 points with an unrestricted encoder, as Proposition~\ref{prop:bounds}(a) requires, and the true mask narrowed them by 25\% at $\epsilon=0.5\,\epsilon^*$ and by about 10\% at larger values.

\begin{table}[htbp]
\centering
\caption{Unsupported histories, strategy ``stop at the landmark'': mechanism-indexed bounds $B(\epsilon,M)$ for $\pout(\mathrm s)$ (true value 5.67), $\omega=1$, 200 replicates.}
\label{tab:bounds}
\small
\begin{tabular}{ccccccc}
\toprule
 & \multicolumn{3}{c}{Unrestricted encoder} & \multicolumn{3}{c}{True mask $M^*$} \\
\cmidrule(lr){2-4}\cmidrule(lr){5-7}
$\epsilon/\epsilon^*$ & Coverage & Width & Empty & Coverage & Width & Empty \\
\midrule
0    & 0.000 & --    & 1.000 & 0.000 & --    & 1.000 \\
0.10 & 0.000 & --    & 1.000 & 0.000 & --    & 1.000 \\
0.25 & 0.000 & --    & 1.000 & 0.000 & --    & 1.000 \\
0.50 & 1.000 & 2.234 & 0.000 & 0.995 & 1.683 & 0.000 \\
1.00 & 1.000 & 2.849 & 0.000 & 1.000 & 2.551 & 0.000 \\
2.00 & 1.000 & 3.016 & 0.000 & 1.000 & 2.740 & 0.000 \\
\bottomrule
\end{tabular}

\smallskip
\parbox{0.92\linewidth}{\footnotesize Coverage, share of replicates in which the estimated set contains the true value (an empty set does not); width in percentage points, averaged over non-empty sets; empty, share of replicates in which no twin met the fit threshold. Mechanism-free iterated regression: bias $-0.49$.}
\end{table}

\paragraph{Back-testing.} The calibrated level was $\hat\epsilon_{\mathrm L}=0.5\,\epsilon^*$ in every replicate, so it never exceeded $\epsilon^*$, as Proposition~\ref{prop:calibration}(a) requires. In this design, however, back-testing added nothing to the fit-based rejection: because the retained intake reduction acts on all patients who stop, the era-1 data already determine its size, and $\hat\epsilon_{\mathrm L}$ coincides with the smallest residual size that the era-1 data do not reject. Back-testing is informative when the omitted mechanism acts mainly on patients for whom the strategy is unsupported, so that a small residual fits the era-1 data but extrapolates wrongly. The present design does not include this case.

\section{Discussion}
\label{sec:discussion}

A digital twin that combines a mechanism with learned components returns an answer for every strategy, and this paper separates the parts of that answer by their source. Where patients followed the strategy, the twin serves as an outcome model inside a doubly robust estimator, and the conclusion does not depend on which of several equally well-fitting twins is used. Where they did not, the twin's answer is reported as a set indexed by how far the twin may depart from the mechanism, and the physics-informed penalty that controls this departure becomes an explicit sensitivity parameter. When practice changes over time, back-testing tells which departures are too small to be consistent with the data.

The simulation illustrates both parts of this division. Four twins that fit the same data disagreed by more than a percentage point of weight loss as plug-in estimators, and the correction brought them to within a few hundredths of a point of each other and of the truth. On patients whom the data did not cover, a mechanism-free extrapolation was off by half a point and gave no sign of it, whereas the bounds were empty when the residual was too small to fit the data and covered the truth once it was not; knowledge constraints on the encoder narrowed them. Two findings qualify these results. The variance estimator of the correction was conservative, so coverage exceeded the nominal level; and in the present design back-testing could not improve on fit-based rejection, because the omitted mechanism also acted on supported patients.

The framework complements our earlier work. \citet{cao2026interference} showed that learned models with the same predictive accuracy can imply different causal conclusions; Theorem~\ref{thm:dr} shows when such disagreement disappears and Proposition~\ref{prop:bounds} quantifies it when it does not. \citet{cao2026prediction} showed that prediction error is a poor guide to causal performance; accordingly, no result here relies on the predictive accuracy of the twin. \citet{cao2026reconstruction} studied when a learned representation is an adequate adjustment set; here the adjustment set is the observed history, and the mechanism enters only through the model for outcome dynamics.

The work has limitations. Sequential exchangeability given the observed history is assumed; reasons for stopping such as cost and insurance coverage may be incompletely recorded, and sensitivity analysis for unmeasured confounding of the decision process is a natural extension. The overlap gap of Assumption~\ref{ass:gap} simplifies the treatment of histories with small but positive decision probabilities. Individual mechanistic parameters may be weakly identified from sparse clinical weights, a difficulty already noted for energy-balance models \citep{hall2024plateau}; the population estimands studied here are less sensitive to this than individual forecasts. Back-testing relies on invariant dynamics across eras, and its coverage guarantee on the remaining unsupported set relies on an assumption that cannot be checked. Finally, the framework has not yet been applied to clinical data.

\section*{Data and Code Availability}

No data were analyzed in this version. The simulation code and instructions for reproducing Section~\ref{sec:simresults} are available at \url{https://github.com/congca0/causal-digital-twins}.

\section*{Funding}

The author received no specific funding for this work.

\section*{Competing Interests}

The author declares no competing interests.

\appendix

\section{Proofs}
\label{app:proofs}

\paragraph{Proposition~\ref{prop:decomp}.} Part (a) is the law of total expectation. For (b), fix $h_0\in\Sset$. By Assumption~\ref{ass:gap}, $\pi_k\ge\delta$ along every continuation reachable under $\bara$, so each $Q_k$ is well defined on those histories. Backward induction using consistency and sequential exchangeability \citep{robins1986} gives $\E(Y^{\bara}\mid \Hist_k=\bar h_k,\bar A_{k-1}=\bar a_{k-1})=Q_k(\bar h_k)$ for reachable $\bar h_k$; at $k=0$ this is $\E(Y^{\bara}\mid H_0=h_0)=Q_0(h_0)$, and averaging over $H_0\in\Sset$ gives the result. For (c), let $\mathcal N=\{h_0\notin\Sset:\pi_0(h_0)=0\}$, which has positive probability. On $\mathcal N$ the observed-data law contains no information on the conditional law of $Y^{\bara}$, because $A_0\ne a$ almost surely. Given any law $P^\dagger$ compatible with the observed data and the assumptions, modify the conditional law of the potential trajectories under $\bara$ given $H_0$ on $\mathcal N$ only, independently of the observed treatments, keeping $Y^{\bara}$ within $[y_{\mathrm{lo}},y_{\mathrm{hi}}]$. The observed-data law and consistency are unaffected, because potential outcomes under $\bara$ are never observed on $\mathcal N$. Exchangeability at $k=0$ is preserved because the modified law does not depend on $A_0$, and exchangeability at $k\ge1$ conditions on $A_0=a$, an event of probability zero on $\mathcal N$. Under the assumption of part (c), $\mathcal N=\Sset^c$. Taking the modified law of $Y^{\bara}$ given $H_0$ on $\mathcal N$ to be a point mass at any $c\in[y_{\mathrm{lo}},y_{\mathrm{hi}}]$ gives $\pout(a)=c$, so every point of the interval is attained; conversely $\pout(a)\in[y_{\mathrm{lo}},y_{\mathrm{hi}}]$ because $Y^{\bara}$ takes values there. Hence the interval is the sharp identified set.

\paragraph{Lemma~\ref{lem:remainder}.} Write $\Delta_k=\hat Q_k-Q_k$, with $\Delta_K=0$, and $\tilde Q_k(\bar h_k)=\E\{\hat Q_{k+1}\mid\Hist_k=\bar h_k,A_k=a\}$, so that $\tilde Q_k-Q_k=\E\{\Delta_{k+1}\mid\Hist_k,A_k=a\}$. All expectations below include the factor $\ind(H_0\in\Sset)$, which is a function of $H_0$ and is omitted from the notation. Because $\hat W_k$ is non-zero only if $A_k=a$, and the nuisance estimates are fixed,
\[
\E\{\hat W_k(\hat Q_{k+1}-\hat Q_k)\}=\E\{\hat W_k(\tilde Q_k-\hat Q_k)\}=\E\Big\{\hat W_{k-1}\frac{\pi_k}{\hat\pi_k}(\tilde Q_k-\hat Q_k)\Big\},
\]
where the second equality integrates over $A_k$ given $\Hist_k$. Writing $\tilde Q_k-\hat Q_k=(\tilde Q_k-Q_k)-\Delta_k$ and using $\E\{\hat W_{k-1}(\pi_k/\hat\pi_k)(\tilde Q_k-Q_k)\}=\E(\hat W_k\Delta_{k+1})$,
\[
\E(\hat\varphi)-\E(Q_0)=\E(\Delta_0)+\sum_{k=0}^{K-1}\E(\hat W_k\Delta_{k+1})-\sum_{k=0}^{K-1}\E\Big(\hat W_{k-1}\frac{\pi_k}{\hat\pi_k}\Delta_k\Big).
\]
Shifting the index in the first sum and using $\Delta_K=0$ and $\hat W_{-1}=1$ gives $\sum_{k=0}^{K-1}\E\{\hat W_{k-1}(1-\pi_k/\hat\pi_k)\Delta_k\}$, which is the stated expression.

\paragraph{Theorem~\ref{thm:dr}.} Write $s=\ind(H_0\in\Sset)$, $\varphi$ for $\hat\varphi$ evaluated at the true nuisances, so that $D_{\mathrm{in}}=s(\varphi-\pin)/p_{\Sset}$, and $\mathbb P$ for expectation over a new observation with the nuisance estimates held fixed. With $J$ folds of sizes $n_j$, the estimator is $\hat\psi_{\mathrm{in}}=\mathbb P_n(s\hat\varphi)/\mathbb P_n(s)$, so
\begin{gather*}
\hat\psi_{\mathrm{in}}-\pin=\frac{\mathbb P_n\{s(\hat\varphi-\pin)\}}{\mathbb P_n(s)},\\
\mathbb P_n\{s(\hat\varphi-\pin)\}=(\mathbb P_n-\mathbb P)\{s(\varphi-\pin)\}+\sum_j\frac{n_j}{n}(\mathbb P_{n,j}-\mathbb P)\{s(\hat\varphi_j-\varphi)\}+\sum_j\frac{n_j}{n}R_j,
\end{gather*}
where $\hat\varphi_j$ uses nuisances fitted outside fold $j$, $\mathbb P_{n,j}$ is the empirical measure of fold $j$, and $R_j=\mathbb P\{s(\hat\varphi_j-\pin)\}=\mathbb P\{s(\hat\varphi_j-Q_0)\}$ because $\E(sQ_0)=p_{\Sset}\pin$; $R_j$ is the remainder of Lemma~\ref{lem:remainder} for the fold-$j$ nuisances. Conditionally on the data outside fold $j$, the middle term has mean zero and variance at most $\|s(\hat\varphi_j-\varphi)\|^2/n_j$, so it is $O_P(\|\hat\varphi_j-\varphi\|/\sqrt n)$ by Chebyshev's inequality \citep{kennedy2016}. Since $Y$ is bounded and $\hat\pi_k,\pi_k\ge\delta/2$ on supported histories, the weights satisfy $\hat W_k\le(2/\delta)^{k+1}$ and $\|\hat\varphi_j-\varphi\|\le C\sum_k(\|\hat Q_k-Q_k\|+\|\hat\pi_k-\pi_k\|)$ for a constant $C$ depending on $\delta$, $K$, and the outcome range; the middle term is therefore $o_P(n^{-1/2})$ under the $L_2$ convergence in (b) and $o_P(1)$ under the convergence to limits in (a). By the Cauchy--Schwarz inequality and $\hat W_{k-1}\le(2/\delta)^{k}$,
\[
|R_j|\le\sum_{k=0}^{K-1}(2/\delta)^{k+1}\,\|\hat\pi_k-\pi_k\|\,\|\hat Q_k-Q_k\|\le C_{\delta,K}\sum_{k=0}^{K-1}\|\hat\pi_k-\pi_k\|\,\|\hat Q_k-Q_k\|,\qquad C_{\delta,K}=(2/\delta)^{K},
\]
with norms in $L_2$ over supported histories along which the strategy has been followed. Finally, $\mathbb P_n(s)=p_{\Sset}+O_P(n^{-1/2})$ and the numerator is $O_P(n^{-1/2})$ in case (b), so the ratio equals the numerator divided by $p_{\Sset}$ up to $o_P(n^{-1/2})$. For (a), each term of the bound converges to $\|\bar\pi_k-\pi_k\|\,\|\bar Q_k-Q_k\|=0$, and the other terms are $o_P(1)$. For (b), $\hat\psi_{\mathrm{in}}-\pin=(\mathbb P_n-\mathbb P)D_{\mathrm{in}}+o_P(n^{-1/2})$ and the central limit theorem applies. Part (c) follows by applying (b) to each procedure, which yields the same leading term $(\mathbb P_n-\mathbb P)D_{\mathrm{in}}$, and, under misspecification of the twins, by applying (a) with $\bar\pi_k=\pi_k$; the plug-in estimator $\mathbb P_n(s\hat Q_0)/\mathbb P_n(s)$ converges to $\E\{\bar Q_0^{(m)}(H_0)\mid H_0\in\Sset\}$.

\paragraph{Proposition~\ref{prop:bounds}.} (a) $\Rset_\epsilon\subseteq\Rset_{\epsilon'}$ and $\mathcal G_M\subseteq\mathcal G_{M'}$ imply $\Tset(\epsilon,M)\subseteq\Tset(\epsilon',M')$, and the infimum and supremum over a larger set are weakly more extreme. (b) Under Assumptions~\ref{ass:seq} and~\ref{ass:adequacy}, $\mathsf T^*$ generates both the potential and the observed trajectories, so it is compatible with $P$; it belongs to $\Tset(\epsilon,M)$ whenever $\epsilon\ge\epsilon^*$ and $\mathcal G_{M^*}\subseteq\mathcal G_M$, and $\pout(\mathsf T^*)=\pout(a)$. (c) At $\epsilon=0$ every twin is determined by its encoder. By assumption, compatible encoders agree almost surely on the support of $X$ among unsupported patients, so they imply the same post-landmark law under $\bara$ for those patients and the same $\pout(\mathsf T)$. If no twin is compatible, the observed data are inconsistent with Assumption~\ref{ass:adequacy} at $\epsilon^*=0$ with mask $M$. (d) Every compatible twin, combined with the observed decision process, defines a law that reproduces $P$ and satisfies Assumptions~\ref{ass:seq} and~\ref{ass:gap}, because the twin generates potential trajectories from $H_0$ without reference to later decisions, so that sequential exchangeability holds by construction. Hence $\pout(\mathsf T)\in\mathcal I_0$ for every compatible twin, and $B(\epsilon,M)\subseteq\overline{\mathcal I_0}$. Conversely, for $c\in\mathcal I_0$ and $\eta>0$, richness gives a compatible twin with finite residual size $\epsilon_{c,\eta}$ and $|\pout(\mathsf T)-c|<\eta$; this twin belongs to $\Tset(\epsilon,M)$ for $\epsilon\ge\epsilon_{c,\eta}$, so $\bigcup_\epsilon B(\epsilon,M)$ meets every neighbourhood of every point of $\mathcal I_0$.

\paragraph{Proposition~\ref{prop:calibration}.} (a) By Assumption~\ref{ass:invariance}, $\mathsf T^*$ generates the potential trajectories in both eras, so it is compatible with $P^{(1)}$ and belongs to the era-1 set at every $\epsilon\ge\epsilon^*$. Under the era-2 identification assumptions, $\psi_{\mathcal V}=\E^{(2)}\{Q_0^{\mathsf T^*}(H_0)\mid H_0\in\mathcal V\}$, which is the value implied by $\mathsf T^*$ and therefore lies in $B^{(1)}_{\mathcal V}(\epsilon^*,M)$. Hence $\epsilon^*$ belongs to the set whose infimum defines $\epsilon_{\mathrm{cal}}$. (b) By monotonicity, $\psi_{\mathcal V}\notin B^{(1)}_{\mathcal V}(\epsilon,M)$ for every $\epsilon<\epsilon_{\mathrm{cal}}$, and by validity this is impossible if Assumption~\ref{ass:adequacy} held at level $\epsilon$. (c) Let $\epsilon_{\mathcal U}$ denote the infimum in the stated condition, so $\epsilon_{\mathcal U}\le\epsilon_{\mathrm{cal}}$. For $\epsilon>\epsilon_{\mathrm{cal}}$ there is $\epsilon'\in[\epsilon_{\mathcal U},\epsilon)$ with $\pout^{\mathcal U}(a)\in B^{(2)}_{\mathcal U}(\epsilon',M)$, and monotonicity gives membership at $\epsilon$. If the infimum $\epsilon_{\mathcal U}$ is attained, the same argument applies at $\epsilon=\epsilon_{\mathrm{cal}}$.

\paragraph{Proposition~\ref{prop:setcons}.} Let $\mathcal L^*=\min_{\Theta_\epsilon}\mathcal L$. On the event $\sup|\mathcal L_n-\mathcal L|\le c_n/2$, whose probability tends to one because $c_n/a_n\to\infty$, every $\theta\in\Theta_0(\epsilon)$ satisfies $\mathcal L_n(\theta)\le\mathcal L^*+c_n/2\le\inf\mathcal L_n+c_n$, so $\Theta_0(\epsilon)\subseteq\hat\Theta(\epsilon)$; and every $\theta\in\hat\Theta(\epsilon)$ satisfies $\mathcal L(\theta)\le\mathcal L^*+2c_n$. By compactness and continuity of $\mathcal L$, for every $\eta>0$ there is $\kappa>0$ such that $\mathcal L(\theta)\le\mathcal L^*+\kappa$ implies $\mathrm{dist}\{\theta,\Theta_0(\epsilon)\}<\eta$; since $c_n\to0$, $\hat\Theta(\epsilon)$ lies in the $\eta$-neighbourhood of $\Theta_0(\epsilon)$ with probability tending to one. This is Hausdorff convergence \citep[compare][Theorem 3.1]{chernozhukov2007}. Uniform continuity of $\pout$ on the compact set and uniform convergence of its sample analogue then give convergence of the endpoints.

\paragraph{Corollary~\ref{cor:lcb}.} At $\epsilon_{j^*}$ the null hypothesis holds by Proposition~\ref{prop:calibration} and monotonicity. By Proposition~\ref{prop:setcons} the estimated bounds at $\epsilon_{j^*}$ converge to their population values, which contain $\psi_{\mathcal V}$ in their interior, and by Theorem~\ref{thm:dr}(b) the Wald interval covers $\psi_{\mathcal V}$ with probability tending to $1-\alpha$ or more. On the intersection of these events the interval intersects the estimated bounds, $H_0(\epsilon_{j^*})$ is not rejected, and $\hat\epsilon_{\mathrm L}\le\epsilon_{j^*}$.

\section{Simulation details}
\label{app:sim}

The simulation is implemented in a single Python script, \texttt{twin\_sim.py}, using NumPy, SciPy, pandas, and scikit-learn. All constants of Section~\ref{sec:simulation} are set at the top of the script, and \texttt{python twin\_sim.py --calibrate} prints the calibration targets reported there. Twins are fitted by nonlinear least squares with the penalty of Remark~\ref{rem:lambda}. For the extremal twins of Algorithm~\ref{alg:extremal}, the constraint $\|r\|_\mu<\epsilon$ is imposed by writing the residual coefficients as $b=\epsilon v/(1+v^\top Gv)^{1/2}$, with $G$ the Gram matrix of Section~\ref{sec:twin}, and each endpoint is found by continuation: an added least-squares term pulls the out-of-support value toward successive targets, and the last target reached within the fit threshold, with relative slack $c_n=2\sqrt{2/N}$, is recorded. Each replicate runs on one core. In the first 50 replicates, with a four-point grid of residual sizes, the median time per replicate was 13 minutes (interquartile range 11--16); the 200 replicates reported in Section~\ref{sec:simresults} used a six-point grid and ran with 30 parallel jobs on a 32-core virtual machine.

\bibliographystyle{plainnat}
\bibliography{references}

\end{document}